\documentclass[nofootinbib,preprint, prX]{revtex4-1} 
\usepackage[utf8]{inputenc}

\usepackage{natbib}
\usepackage{cancel}
\usepackage{xspace}
\usepackage{amsmath}
\usepackage{amsfonts}
\usepackage{graphicx}
\usepackage{hyperref}

\newcommand{\Dt}{\widetilde{D}}
\newcommand{\EFT}{EFT($\cancel{\pi}$)\xspace}

\begin{document}

\title{Effective field theory analysis of boson-trimer bond lengths to next-to-leading order}
\author{Julia Qin}
\email{jq676@nyu.edu}
\affiliation{Department of Physics,
Stetson University,
DeLand, FL 32723
}
\affiliation{New York University, New York, NY 70 Washington Square South 10012}
\author{Jared Vanasse}
\email{jvanass3@fitchburgstate.edu}
\affiliation{Department of Physics,
Stetson University,
DeLand, FL 32723
}
\affiliation{Fitchburg State University,
Fitchburg, MA 160 Pearl St. 01420
}

\begin{abstract}
Cold Helium atoms are a unique system where a single excited three-body Efimov state occurs, naturally, without the need for an external magnetic field. While three-body bound state energies of cold Helium atoms have previously been investigated, recent experimental techniques have allowed their structure to also be studied. The weak interaction between Helium atoms leads to a helium-helium (dimer) scattering length, $a$, much larger than the helium-helium effective range of interaction, $r$.  This feature is exploited in a theory that systematically expands observables in powers of $r/a$, known as short range effective field theory (srEFT), which has been used successfully to investigate properties of cold atom systems.  Using srEFT we investigate the average bond length of atoms in the three-body ground state and excited Efimov state of cold Helium atoms.  At leading-order (next-to-leading order) in srEFT, we find the average bond length of the $^4$He trimer ground state is 8.35(33) $\mathrm{\AA}$ (10.29(2) $\mathrm{\AA}$) and the average bond length of the excited $^4$He trimer Efimov state is 103(4) $\mathrm{\AA}$ (105.3(2) $\mathrm{\AA}$).
\end{abstract}

\date{\today}

\maketitle

\section{Introduction}

In the limit the scattering length between two particles $a\to\infty$, the two-body system (dimer) becomes very weakly bound.  Vitaly Efimov showed that in this limit, the three-body system forms an infinite tower of geometrically spaced three-body bound states (trimers)~\cite{Efimov:1970zz}.  For identical bosons, the ratio between adjacent energy levels is 515. Discovery of the Efimov effect spurred searches for Efimov states in nuclear systems.  However, no real Efimov state has been found in a nuclear system to date~\cite{Jensen:2004zz}.  The first experimental observation  of an Efimov state was in ultracold Cesium~\cite{kraemer2006evidence}, where a magnetic field created a Feschbach resonance~\cite{chin2010feshbach} to control the scattering length between Cesium atoms. The signature of the Efimov state was seen in the loss rate of Cesium atoms near an Efimov resonance.  This Efimov resonance occurred at a negative scattering length between Cesium atoms where the first Efimov trimer becomes bound. Efimov states were also observed in cold Lithium atoms, tuned with a Feschbach resonance and detected using loss measurements~\cite{PhysRevLett.103.130404} and spectroscopy~\cite{Lompe940,PhysRevLett.106.143201}. Cesium atoms were also used to discover the second excited Efimov state~\cite{PhysRevLett.112.190401}. The first observation of Efimov states in cold atoms without the need of a Feshbach resonance was in cold Helium atoms~\cite{Kunitski:2015qth}. Weak van der Waals interactions between Helium atoms causes Helium atoms to form weakly bound dimers and a single excited Efimov state, circumventing the need for a Feshbach resonance to create an Efimov state. The Efimov state in Helium atoms has been imaged using the Coulomb explosion technique, which maps the size and physical structure of Helium trimers~\cite{Kunitski:2015qth}.  Matter wave diffraction has also been used to measure the size of the Helium trimer ground state~\cite{bruhl2005matter}.

Phenomenological potential models have been used to calculate the Helium trimer bound state energy~\cite{blume2000comparative,doi:10.1063/1.4892564,ROUDNEV200097,Kolganova:2011uc,Blume:2015nla,Barletta:2001za} and structural properties~\cite{blume2000comparative,doi:10.1063/1.4892564,ROUDNEV200097,Barletta:2001za,Blume:2015nla}, such as average bond length. Effective field theory (EFT) offers a model independent approach for determining the properties of Helium trimers and estimating errors in theoretical calculations.  When probing distance scales much larger than the range of the underlying interaction, $r$, we can approximate a short range interaction by a series of contact interactions. At long distance scales (or low energies), the short distance details of the interaction are encoded into the coefficients of the contact interactions. An EFT approach systematically expands observables in a ratio of disparate scales. Because the weakly bound Helium dimer has a scattering length $a$ much larger than the effective range $r$, we can use the expansion parameter $r/a$. At leading-order (LO) srEFT has a single two-body parameter and three-body parameter.  The two-body parameter is typically fixed to reproduce the dimer binding energy or the atom-atom scattering length.  A LO three-body force is necessary because the solution to the three-boson problem does not possess a unique solution and needs a three-body counterterm to fix the trimer spectrum and give a unique solution~\cite{Bedaque:1998kg,Bedaque:1998km}.   The loss rate of cold atoms has been calculated in srEFT at LO~\cite{Bedaque:2000ft,Braaten:2001hf,Braaten:2006vd} and next-to-leading order (NLO)~\cite{Ji:2012nj}.  srEFT has calculated the binding energies of Helium trimers and the atom-dimer scattering amplitude to next-to-next-to-leading order (NNLO)~\cite{Platter:2006ev,Ji:2012nj}. Although the bond length of Helium trimers has previously been determined using potential models, the bond length has not yet been calculated in srEFT away from the unitary limit~\cite{Braaten:2004rn}.  This work presents a srEFT calculation of the bond length between Helium atoms in the three-body ground state and excited Efimov state.

Any system described by short ranged interactions possessing relatively shallow bound states can be described systematically using srEFT.  In the context of halo-nuclei this srEFT is known as halo-EFT and in few-nucleon systems as pionless EFT (\EFT). Halo-EFT and \EFT have been used to great success in describing properties of halo-nuclei~(See Ref.~\cite{Hammer:2017tjm}) and few-nucleon systems (See Refs.~\cite{Beane:2000fx,Vanasse:2016jtc}).  The charge radii and matter radii of halo nuclei were calculated in Ref.~\cite{Canham:2008jd,Acharya:2013aea,Hagen:2013xga} and range corrections were considered in~\cite{Canham:2009xg,Vanasse:2016hgn}.  Similar techniques were used to calculate charge radii of three-nucleon systems~\cite{Vanasse:2015fph,Konig:2019xxk,Kirscher:2017fqc} using \EFT.  This work adapts the techniques of Refs.~\cite{Vanasse:2015fph,Vanasse:2016jtc} for halo-nuclei and three-nucleon systems to cold atom trimers.

This paper is organized as follows. Sec.~\ref{sec:srEFT} reviews srEFT. In Sec.~\ref{sec:Three-boson Bound states}, how to perform three-body calculations in srEFT is shown. Sec.~\ref{sec:Size of trimer} calculates the form factor for the Helium trimer up to NLO. Results for the trimer bond length are presented and compared to experimental and theoretical predictions in Sec.~\ref{sec:Results}. Conclusions are given in Sec.~\ref{sec:Conclusion}.

\section{srEFT\label{sec:srEFT}}

At low energies, $S$-wave scattering dominates the scattering between two cold bosons. $S$-wave scattering has the scattering amplitude
\begin{equation}
    \mathcal{A}(k)=\frac{4\pi}{m}\frac{1}{k\cot\delta_0-ik},
\end{equation}
where $k$ is the center-of-mass (c.m.) momentum and $\delta_0$ is the $S$-wave phase shift.  Using the effective range expansion (ERE)~\cite{PhysRev.75.312,PhysRev.75.1637,Bethe:1949yr}, the phase shift contribution can be written as
\begin{equation}
    k\cot\delta_0=-\frac{1}{a}+\frac{1}{2}rk^2-sk^4+\cdots,
\end{equation}
where $a$ is the scattering length, $r$ the effective range of interactions, and $s$ the shape parameter. At low energies, only the first few terms are relevant in the ERE.  A more convenient parametrization of the ERE uses the fact that a pole at positive imaginary values of $k$ corresponds to a boson-boson bound state (dimer).  Expanding around this bound state, the ERE gives
\begin{equation}
    \label{eq:ERE}
    k\cot\delta_0=-\gamma+\frac{1}{2}\rho(\gamma^2+k^2)-\rho_1(\gamma^2+k^2)^2+\cdots,
\end{equation}
where the bound state energy of the dimer state is $B_d=\frac{\gamma^2}{m}$.  Matching coefficients of powers of momentum $k$ these different parametrizations can be matched leading to
\begin{equation}
    \frac{1}{2}r=\frac{1}{2}\rho-2\rho_1\gamma^2,
\end{equation}
\begin{equation}
    \label{eq:ERErho}
    -\frac{1}{a}=-\gamma+\frac{1}{2}\rho\gamma^2+\cdots.
\end{equation}
Because the $\rho_1$ term is suppressed relative to $r$ and $a$, we drop the $\rho_1$ term to find $\rho=r$.  Throughout this work, we will use the ERE expansion about the bound state pole of the dimer.

\begin{figure}[h!]
\centering
\includegraphics[width=16cm]{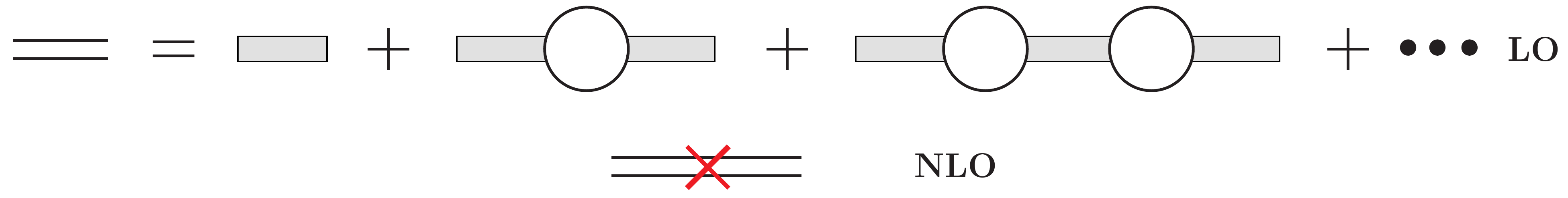}
\caption{\label{fig:dimer}Diagrammatic representation of the dimer propagator at LO and NLO.  The grey bar is the bare dimer propagator given by $i/\Delta$.  Single lines represent a single boson and the double line the dressed dimer propagator.  Finally, the cross represents an insertion of an effective range correction.}
\end{figure}

The Lagrangian in srEFT for interacting bosons is
\begin{align}
\mathcal{L}=&\hat{\psi}^\dagger\left(i\partial_0+\frac{\nabla^2}{2m}\right)\hat{\psi}- \hat{d}^\dagger\left(c_2\left(i\partial_0+\frac{\nabla^2}{4m}+\frac{\gamma^2}{m}\right)-\Delta\right)\hat{d}\\\nonumber&-y\left(\hat{d}^\dagger\hat{\psi}\hat{\psi}+\mathrm{H.c.}\right)+h\hat{d}^\dagger\hat{d}\hat{\psi}^\dagger\hat{\psi},
\end{align}
where $\hat{\psi}$ and $\hat{d}$ are the boson and dimer field, respectively. $y$ sets the strength of the interaction between the dimer and boson, while $h$ is a three-body force term necessary to properly renormalize the three-boson system~~\cite{Bedaque:1998kg,Bedaque:1998km}.  Rather than refitting $h$ at each order we expand it perturbatively as
\begin{equation}
    h=h_{\mathrm{LO}}+h_{\mathrm{NLO}}+\cdots.
\end{equation}
The dressed dimer propagator up to NLO is given by the infinite sum of diagrams in Fig~\ref{fig:dimer}.
The expression for the dressed dimer propagator up to NLO is
\begin{equation}
    i\bar{D}(p_0,\vec{p}\,)=\frac{4\pi}{my^2}\frac{i}{\frac{4\pi}{my^2}\Delta-\sqrt{\frac{\vec{p}^{\,2}}{4}-mp_0-i\epsilon}+\frac{\Lambda}{2\pi}}\left[1+\frac{\frac{c_2}{m}\left(mp_0-\frac{\vec{p}^{\,2}}{4}+\gamma^2\right)}{\frac{4\pi}{my^2}\Delta-\sqrt{\frac{\vec{p}^{\,2}}{4}-mp_0-i\epsilon}+\frac{\Lambda}{2\pi}}\right].
\end{equation}
When calculating loop integrals, we use a cutoff regularization scheme that regulates divergences by dropping all momenta greater than $\Lambda$.  Cutoff regularization introduces additional terms that scale like $1/\Lambda^n$.  These terms are ignored in the analytical expression of the dimer propagator and three-boson numerical computations by taking $\Lambda$ sufficiently large. The boson-boson scattering amplitude in the c.m. frame is related to the dimer propagator via
\begin{equation}
    \mathcal{A}(k)=-y^2\bar{D}\left(\frac{k^2}{m},0\right)=-\frac{4\pi}{m}\frac{1}{\frac{4\pi}{my^2}\Delta+\frac{\Lambda}{2\pi}+ik}\left[1+\frac{\frac{c_2}{m}(\gamma^2+k^2)}{\frac{4\pi}{my^2}\Delta+\frac{\Lambda}{2\pi}+ik}\right].
\end{equation}
Matching this to the ERE, Eq.~(\ref{eq:ERE}), the parameters in the Lagrangian are
\begin{equation}
    y^2=\frac{4\pi}{m},\quad \Delta=\gamma-\frac{\Lambda}{2\pi},\quad c_2=m\frac{1}{2}r.
\end{equation}
The dimer wavefunction renormalization is given by the residue about the dimer bound state pole, which at LO gives
\begin{equation}
    Z_d=\frac{2\gamma}{m}.
\end{equation}

An alternative formalism for the Lagrangian introduces a trimer field $\hat{t}$~\cite{Bedaque:2002yg} and is given by
\begin{align}
    \mathcal{L}=&\hat{\psi}^\dagger\left(i\partial_0+\frac{\nabla^2}{2m}\right)\hat{\psi}+\hat{d}^\dagger\left(c_2\left(i\partial_0+\frac{\nabla^2}{4m}+\frac{\gamma^2}{m}\right)-\Delta\right)\hat{d}+\hat{t}^\dagger\Omega\hat{t}\\\nonumber
    &-y\left(\hat{d}^\dagger\hat{\psi}\hat{\psi}+\mathrm{H.c.}\right)-\omega\left(\hat{t}^\dagger \hat{d}\hat{\psi}+\mathrm{H.c}\right).
\end{align}
Parameters in the two Lagrangians can be matched by performing Gau{\ss}ian integration over the trimer field or by performing a simple matching calculation, yielding
\begin{equation}
    h=-\frac{\omega^2}{\Omega}.
\end{equation}
Introduction of a trimer field eases some features of three-boson bound state calculations~\cite{Vanasse:2015fph}.

\section{Three-boson Bound states\label{sec:Three-boson Bound states}}

\begin{figure}[h!]
\centering
\includegraphics[width=10cm]{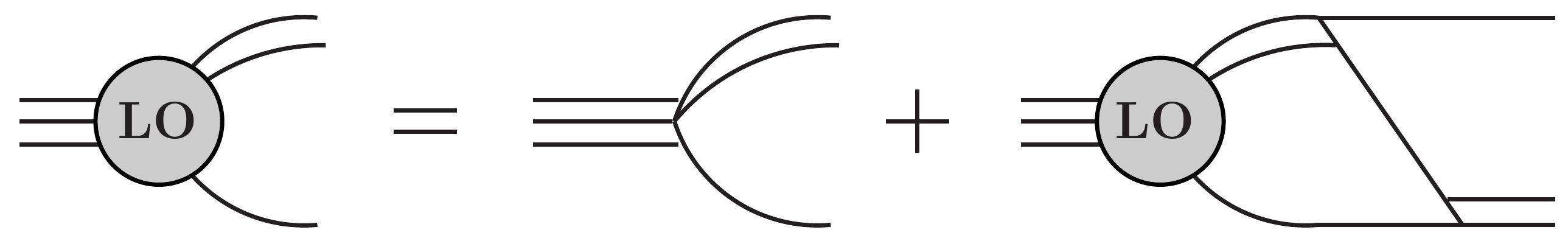}
\caption{\label{fig:vertexLO}Diagrammatic representation of the integral equation for the LO trimer vertex function.  Triple lines represent a trimer propagator and the gray circle the LO trimer vertex function.}
\end{figure}

In order to find properties of the trimer, it is necessary to determine the trimer wavefunction or, equivalently, the trimer vertex function. The LO trimer vertex function is given by an infinite sum of diagrams, which is equivalent to the solution of an integral equation shown diagrammatically in Fig.~\ref{fig:vertexLO}.
This integral equation is given by
\begin{equation}
    \mathcal{G}_0(E,p)=1+K(q,p,E)D(q,E)\otimes_q \mathcal{G}_0(E,q),
\end{equation}
where $\mathcal{G}_0(E,p)$ is the LO trimer vertex function and the subscript ``$0$" denotes it is LO. Projecting the propagator of the exchanged boson in Fig.~\ref{fig:vertexLO} onto the relative $S$-wave state between boson and dimer gives 
\begin{equation}
    K(q,p,E)=\frac{4\pi}{qp}\ln\left(\frac{q^2+p^2+qp-mE-i\epsilon}{q^2+p^2-qp-mE-i\epsilon}\right).
\end{equation}
$D(q,E)$ is the dimer propagator given by
\begin{equation}
    D(q,E)=\frac{1}{\sqrt{\frac{3}{4}q^2-mE-i\epsilon}-\gamma}.
\end{equation}
$\otimes_q$ is shorthand for integration defined by
\begin{equation}
    A(q)\otimes_qB(q)=\frac{1}{2\pi^2}\int_{0}^{\Lambda}dqq^2A(q)B(q).
\end{equation}

\begin{figure}[hbt!]
\centering
\includegraphics[width=11cm]{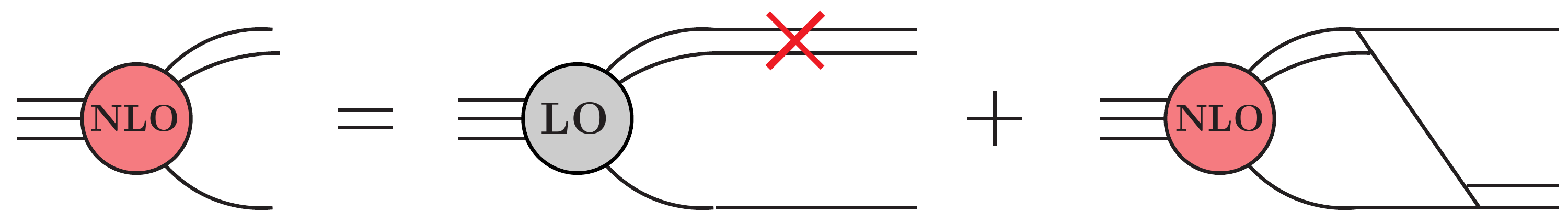}
\caption{\label{fig:vertexNLO}Diagrammatic representation of the integral equation for the NLO correction to the trimer vertex function.}
\end{figure}

At NLO the trimer vertex function receives a range correction giving the NLO correction to the trimer vertex function, which can be solved by using an integral equation shown diagrammatically in Fig.~\ref{fig:vertexNLO}.  
The NLO correction to the trimer vertex function is given by
\begin{equation}
    \mathcal{G}_1(E,p)=R_1(E,p)\mathcal{G}_0(p)+K(q,p,E)D(q,E)\otimes_q \mathcal{G}_1(E,q),
\end{equation}
where the kernel of the integral equation is the same as the LO trimer vertex function and $R_1(E,p)$ containing the range correction is given by
\begin{equation}
    R_1(E,p)=\frac{1}{2}r\left(\gamma+\sqrt{\frac{3}{4}p^2-mE-i\epsilon}\right).
\end{equation}
Finally, the trimer wavefunction renormalization up to NLO is
\begin{equation}
    Z_t=\frac{\pi}{\Sigma_{0}'(B_0)}\left[\underbrace{\vphantom{\frac{\Sigma_1'(B)}{\Sigma_0'(B)}}1}_{\mathrm{LO}}-\underbrace{\frac{\Sigma_1'(B_0)}{\Sigma_0'(B_0)}}_{\mathrm{NLO}}+\cdots\right],
\end{equation}
where $B_0$  is the LO trimer binding energy.  Functions $\Sigma_n(E)$ are given by
\begin{equation}
    \Sigma_{n}(E)=\pi D(E,q)\otimes_q\mathcal{G}_n(E,q).
\end{equation}

\begin{figure}[hbt!]
\centering
\includegraphics[width=11cm]{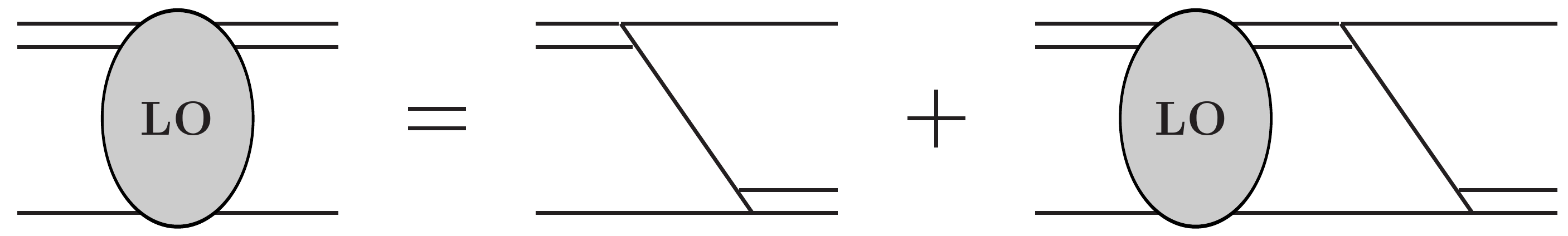}
\caption{\label{fig:LOscatt}Diagrammatic representation of the integral equation for the LO boson-dimer scattering amplitude with no three-body forces.}
\end{figure}

\begin{figure}[hbt!]
\centering
\includegraphics[width=11cm]{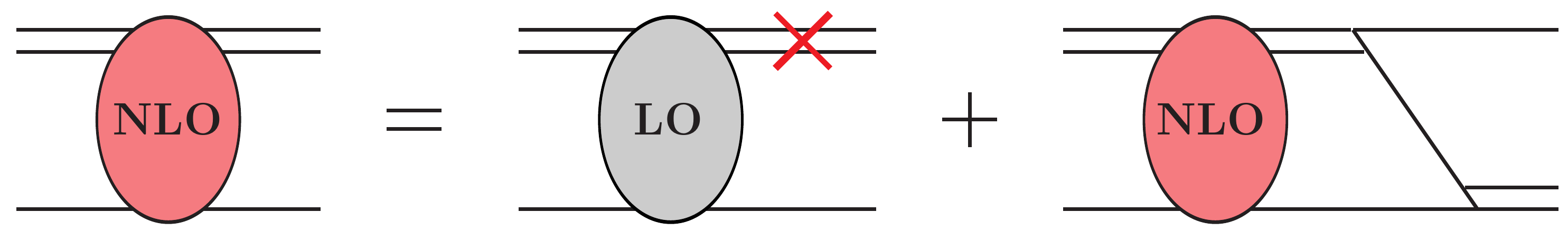}
\caption{\label{fig:NLOscatt}Diagrammatic representation of the integral equation for the NLO correction to the boson-dimer scattering amplitude with no three-body forces.}
\end{figure}

The LO boson-dimer scattering amplitude without three-body forces is given by an infinite sum of diagrams that can be calculated via an integral equation represented diagrammatically in Fig~\ref{fig:LOscatt}
and is given by 
\begin{equation}
    t_0(k,p,E)=B_0(k,p,E)+K(q,p,E)D(q,E)\otimes_q t_0(k,q,E),
\end{equation}
where 
\begin{equation}
    B_0(k,p,E)=\frac{1}{kp}\ln\left(\frac{k^2+p^2+kp-mE-i\epsilon}{k^2+p^2-kp-mE-i\epsilon}\right).
\end{equation}
$k$ ($p$) is the incoming (outgoing) c.m.~momentum between the boson and dimer.  The outgoing boson and dimer are off-shell but the incoming boson and dimer are on-shell giving the condition $E=\frac{3k^2}{4m}-B_d$.  The NLO correction to the boson-dimer scattering amplitude without three-body forces is given by the integral equation
\begin{equation}
    t_1(k,p,E)=R_1(E,p)t_0(k,p,E)+K(q,p,E)D(q,E)\otimes_q t_1(k,q,E),
\end{equation}
represented diagrammatically in Fig.~\ref{fig:NLOscatt}. 

With the LO scattering amplitude without three-body forces and its NLO correction the LO three-body force can be fit to the dimer-boson scattering length, $a_{ad}$, via
\begin{equation}
    h_{\mathrm{LO}}=\frac{x}{1+x\Sigma_0(-B_d)},
\end{equation}
where
\begin{equation}
    x=\frac{-\left(\frac{3\pi a_{ad}}{m}+Z_dt_0(0,0,-B_d)\right)}{\pi Z_d\left[\mathcal{G}_0(-B_d,0)\right]^2}.
\end{equation}
From the LO three-body force the LO binding energy of the trimer can be found through the condition $\Delta_3(B_0)=0$, where
\begin{equation}
    \Delta_3(B)=1-h_{\mathrm{LO}}\Sigma_0(B).
\end{equation}
The NLO correction to the trimer binding energy, $B_1$, is given by
\begin{equation}
    B_1=-\frac{h_{\mathrm{LO}}\Sigma_1(B_0)+h_{\mathrm{NLO}}\Sigma_0(B_0)}{h_{\mathrm{LO}}\Sigma_0'(B_0)},
\end{equation}
where the NLO correction to the three-body force, $h_{\mathrm{NLO}}$, is given by
\begin{align}
   & h_{\mathrm{NLO}}=-\Bigg{\{}\Big{[}t_1(0,0,-B_d)+\gamma rt_0(0,0,-B_d)\Big{]}\Delta_3^2(B_0)+\pi h_{\mathrm{LO}}^2\Sigma_1(B_0)[\mathcal{G}_0(-B_d,0)]^2\\\nonumber
    &\hspace{3cm}+\pi \Delta_3(B_0)\mathcal{G}_0(-B_d,0)\Big{[}h_{\mathrm{LO}}\mathcal{G}_0(-B_d,0)+2h_{\mathrm{LO}}\mathcal{G}_1(-B_d,0)\Big{]}\Bigg{\}}\\\nonumber
    &\hspace{3cm}\Bigg{\{}\pi h_{\mathrm{LO}}\Sigma_0(B_0)\left[\mathcal{G}_0(-B_d,0)\right]^2\Bigg{\}}^{-1}.
\end{align}
For details of how these three-body forces and bound state energies are determined see Ref.~\cite{Vanasse:2015fph}.

\section{Size of trimer\label{sec:Size of trimer}}
\begin{figure}[h!]
\centering
\includegraphics[width=10cm]{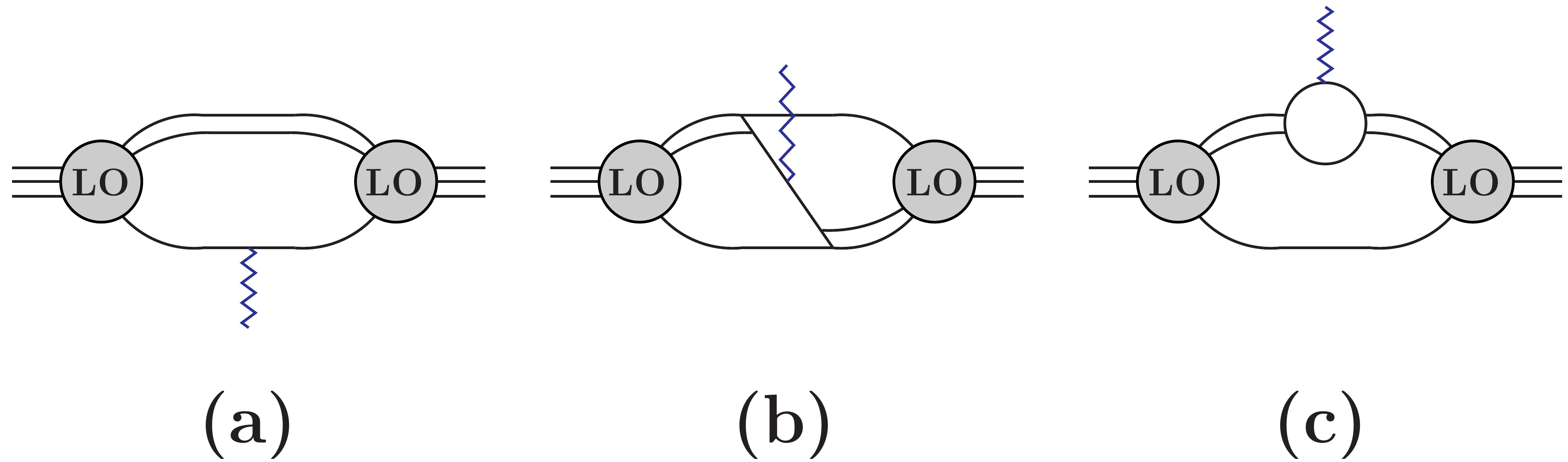}
\caption{\label{fig:LOFF}Diagrams for the LO form factor.}
\end{figure}

\begin{figure}[h!]
\centering
\includegraphics[width=10cm]{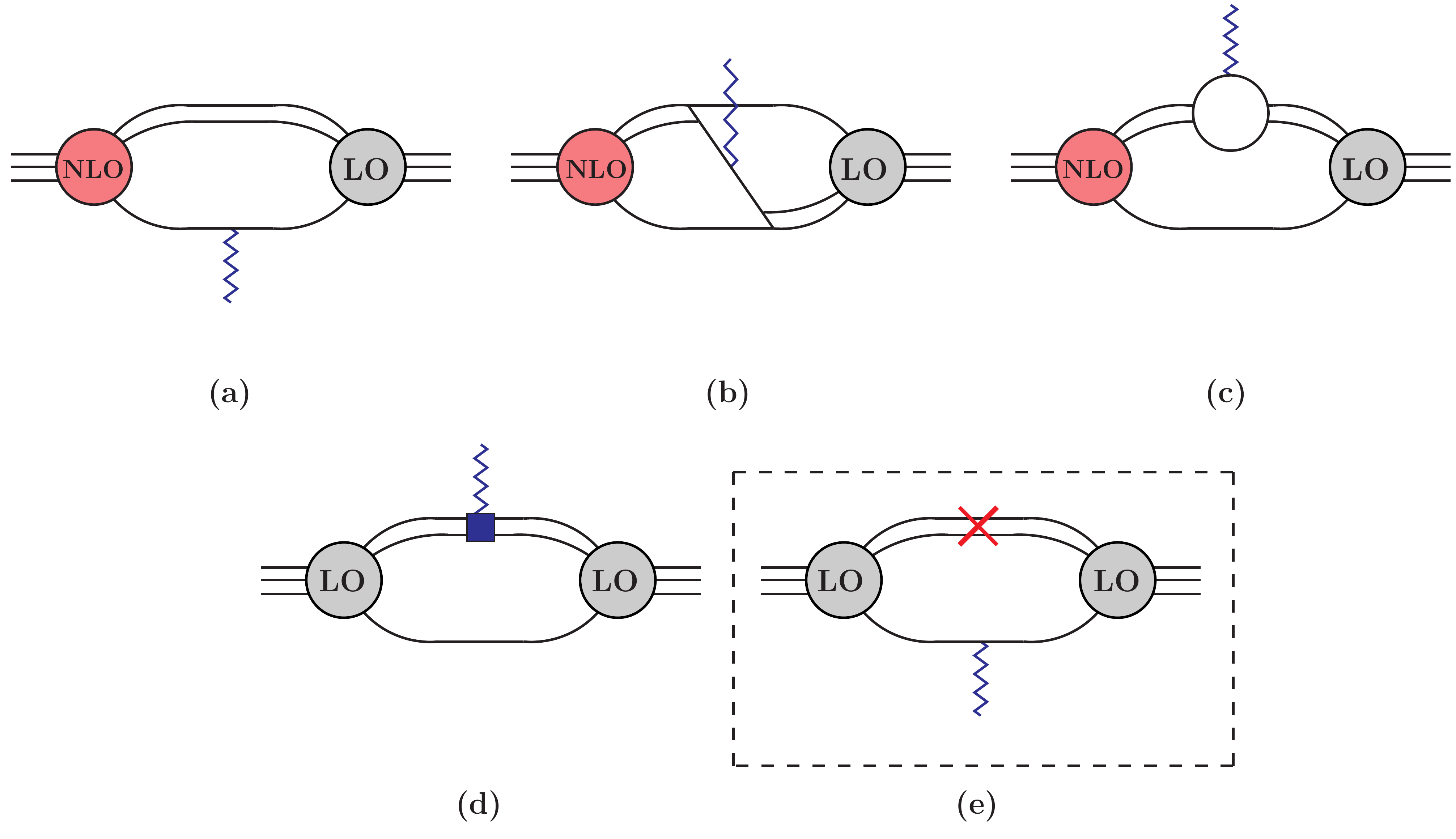}
\caption{\label{fig:NLOFF}Diagrams for the NLO correction to the form factor.  Diagrams related by time reversal symmetry are not shown.  The boxed diagram is subtracted to avoid double counting from diagram (a) and its time reversed version.}
\end{figure}

One approach to determine the size of the trimer is to calculate the form factor of an interacting probe.  A simple way to do this is to gauge the derivatives in the Lagrangian giving bosons an effective charge.  Carrying out this procedure the LO form factor is given by the sum of diagrams in Fig.~\ref{fig:LOFF}, 
where the zig-zag line represent an insertion of a single current probe from gauging the time derivatives of the boson Lagrangian.  Choosing the so called Breit frame in which the probe has zero energy and momentum $\vec{Q}$, the resulting form factor, $F_0(Q^2)$ is a function of $Q^2$ and can be expanded in powers of $Q^2$ yielding
\begin{equation}
    F_0(Q^2)=1-\frac{1}{6}\langle r^2\rangle_0 Q^2+\cdots.
\end{equation}
The LO form factor is normalized at $Q^2=0$ such that it equals one, and the coefficient of the $Q^2$ contribution to the form factor is proportional to the average squared radius of bosons from the c.m.~of the trimer state.  In principle the form factor can be calculated for various small powers of $Q$ and a polynomial can be fit to find the radius squared of the trimer state.  An alternative approach, we take here, is to directly calculate the constant term and $Q^2$ contribution to the form factor by expanding the analytical expression for the form factor in powers of $Q^2$ and picking out each respective piece.  The constant term of the LO form factor is given by
\begin{equation}
    1=2\pi m Z_{t}^{\mathrm{LO}}\widetilde{\mathcal{G}}_0(q)\otimes_q\left\{\frac{\pi}{2}\frac{\delta(q-\ell)}{q^2\sqrt{\frac{3}{4}q^2-mB_0}}-\frac{4}{q^2\ell^2-(q^2+\ell^2-mB_0)^2}\right\}\otimes_\ell\widetilde{\mathcal{G}}_0(\ell),
\end{equation}
where
\begin{equation}
    \widetilde{\mathcal{G}}_n(p)=D(p,E)\mathcal{G}_n(B_0,p).
\end{equation}
The NLO form factor is given by the sum of diagrams in Fig.~\ref{fig:NLOFF}
where diagrams with the NLO vertex function on the right are not shown.  Diagram (e) is subtracted from the other diagrams to avoid double counting from diagram (a) and the time reversed version with the NLO vertex function on the right hand side.  Diagram (d) comes from gauging the time derivative acting on the dimer.  This term is necessary to ensure that the NLO correction to the form factor does not change the form factor normalization at $Q^2=0$.  The NLO correction to the form factor has the $Q^2$ expansion
\begin{equation}
    F_1(Q^2)=-\frac{1}{6}\langle r^2\rangle_1 Q^2+\cdots
\end{equation}
where $\langle r^2\rangle_1$ is the NLO correction to the squared radius of the trimer.  Since the LO form factor at $Q^2=0$ is already normalized to one the constant piece of the NLO correction to the form factor must give zero leading to the condition
\begin{align}
    0=&4\pi m Z_{t}^{\mathrm{LO}}\left[\widetilde{\mathcal{G}}_1(q)-\frac{\Sigma_1'(B_0)}{\Sigma_0'(B_0)}\widetilde{\mathcal{G}}_0(q)\right]\otimes_q\left\{\frac{\pi}{2}\frac{\delta(q-\ell)}{q^2\sqrt{\frac{3}{4}q^2-mB_0}}\right.\\[5mm]\nonumber
    &\hspace{.5cm}\left.-\vphantom{\frac{\pi}{2}\frac{\delta(q-\ell)}{q^2\sqrt{\frac{3}{4}q^2-mB_0}}}\frac{4}{q^2\ell^2-(q^2+\ell^2-mB_0)^2}\right\}\otimes_\ell\widetilde{\mathcal{G}}_0(\ell)-2\pi mr Z_{t}^{\mathrm{LO}}\widetilde{\mathcal{G}}_0(q)\otimes_q\left\{\frac{\pi}{2}\frac{\delta(q-\ell)}{q^2}\right\}\otimes_{\ell}\widetilde{\mathcal{G}}_0(\ell).
\end{align}

To find the size of the trimer it is necessary to obtain the coefficient of the $Q^2$ contribution of the form factor.  The coefficient of the $Q^2$ contribution from type (a) diagrams at LO and NLO to the form factor is given by
\begin{align}
    \label{eq:FFQ2A}
    &\frac{1}{2}\frac{\partial^2}{\partial Q^2}F_{n}^{(a)}(Q^2)=Z_{t}^{\mathrm{LO}}\sum_{i,j=0}^{i+j\leq n}\left\{\tilde{\mathcal{G}}_i(p)\otimes_p\mathcal{A}_{n-i-j}(p,k)\otimes_k\tilde{\mathcal{G}}_j(k)\right.\\\nonumber
    &\hspace{5cm}\left.+2\tilde{\mathcal{G}}_i(p)\otimes_p\mathcal{A}_{n-i}(p)\delta_{j0}+\mathcal{A}_n\delta_{i0}\delta_{j0}\right\}.
\end{align}
Functions $\mathcal{A}_n(p,k)$, $\mathcal{A}_n(p)$, and $\mathcal{A}_n$, are given in Appendix~\ref{sec:FFQ2}. $Q^2$ contributions from diagram (b) at LO and NLO are given by
\begin{align}
    \label{eq:FFQ2B}
    &\frac{1}{2}\frac{\partial^2}{\partial Q^2}F_{n}^{(b)}(Q^2)=Z_{t}^{\mathrm{LO}}\sum_{i=0}^{ n}\tilde{\mathcal{G}}_i(p)\otimes_p\mathcal{B}_{0}(p,k)\otimes\tilde{\mathcal{G}}_{n-i}(k),
\end{align}
and the form of $\mathcal{B}_0(p,k)$ is in Appendix~\ref{sec:FFQ2}.  Contributions to $Q^2$ from type (c) diagrams at LO and NLO are
\begin{align}
    \label{eq:FFQ2C}
    &\frac{1}{2}\frac{\partial^2}{\partial Q^2}F_{n}^{(c)}(Q^2)=Z_{t}^{\mathrm{LO}}\sum_{i,j=0}^{i+j\leq n}\left\{\tilde{\mathcal{G}}_i(p)\otimes_p\mathcal{C}_{n-i-j}(p,k)\otimes_k\tilde{\mathcal{G}}_j(k)\right.\\\nonumber
    &\hspace{5cm}\left.+2\tilde{\mathcal{G}}_i(p)\otimes_p\mathcal{C}_{n-i}(p)\delta_{j0}\right\},
\end{align}
where again $\mathcal{C}_n(p,k)$ and $\mathcal{C}_n(p)$ are given in Appendix~\ref{sec:FFQ2}.  Finally, at NLO the $Q^2$ contribution from diagram (d) is
\begin{align}
    \label{eq:FFQ2D}
    &\frac{1}{2}\frac{\partial^2}{\partial Q^2}F_{1}^{(d)}(Q^2)=Z_{\psi}^{\mathrm{LO}}\left\{\tilde{\mathcal{G}}_0(p)\otimes_p\mathfrak{D}_{1}(p,k)\otimes_k\tilde{\mathcal{G}}_0(k)+2\tilde{\mathcal{G}}_0(p)\otimes_p\mathfrak{D}_{1}(p)\right\},
\end{align}
where the functions $\mathfrak{D}_1(p,k)$ and $\mathfrak{D}_1(p)$ are in Appendix~\ref{sec:FFQ2}.  The LO $Q^2$ contribution to the form factor is given by the sum of contributions from diagrams (a), (b), and (c) in Fig.~\ref{fig:LOFF} yielding
\begin{equation}
    \frac{1}{2}\frac{\partial^2}{\partial Q^2}F_0(Q^2)\Big{|}_{Q^2=0}=\frac{1}{2}\frac{\partial^2}{\partial Q^2}\left(F_0^{(a)}(Q^2)+F_0^{(b)}(Q^2)+F_0^{(c)}(Q^2)\right)\Big{|}_{Q^2=0}.
\end{equation}
At NLO the $Q^2$ contribution of the form factor is given by the contributions from the NLO diagrams (a), (b), (c), and (d) in Fig.~\ref{fig:NLOFF} and the LO form factor times the NLO correction to the trimer wavefunction renormalization which yields
\begin{align}
    \frac{1}{2}\frac{\partial^2}{\partial Q^2}F_0(Q^2)\Big{|}_{Q^2=0}&=\frac{1}{2}\frac{\partial^2}{\partial Q^2}\left(\vphantom{\frac{\Sigma_1'(B_0)}{\Sigma_0'(B_0)}}F_1^{(a)}(Q^2)+F_1^{(b)}(Q^2)+F_1^{(c)}(Q^2)+F_1^{(d)}(Q^2)\right.\\\nonumber
    &\hspace{7cm}\left.-\frac{\Sigma_1'(B_0)}{\Sigma_0'(B_0)}F_0(Q^2)\right)\Big{|}_{Q^2=0}.
\end{align}

\section{Results\label{sec:Results}}
The LO (NLO correction to the) radius squared $\langle r^2\rangle_0$ ($\langle r^2\rangle_1$) of the trimer state is obtained from the form factor by
\begin{equation}
    \langle  r^2\rangle_n=-6\left(\frac{1}{2}\frac{\partial^2}{\partial Q^2}F_n(Q^2)\Big{|}_{Q^2=0}\right).
\end{equation}
This trimer radius assumes that the Helium atoms with respect to the probe are point particles.  The trimer radius up to NLO is given by
\begin{equation}
     r=\sqrt{\langle r^2\rangle_0}\left(\underbrace{\vphantom{\frac{1}{2}\frac{\langle r^2\rangle_1}{\langle r^2\rangle_0}+\frac{1}{2}\frac{\frac{\partial\langle r^2\rangle_1}{\partial E}\Big{|}_{E=B_0}}{\langle r^2\rangle_0}B_1}1}_{\mathrm{LO}}+\underbrace{\frac{1}{2}\frac{\langle r^2\rangle_1}{\langle r^2\rangle_0}+\frac{1}{2}\frac{\frac{\partial\langle r^2\rangle_0}{\partial E}\Big{|}_{E=B_0}}{\langle r^2\rangle_0}B_1}_{\mathrm{NLO}}\right),
\end{equation}
where the final term is due to a NLO correction, $B_1$, to the trimer binding energy.  Although the derivative of the NLO trimer radius squared can be calculated analytically we calculate it numerically.  If the binding energy of the trimer is fixed at each order the expression for the trimer radius reduces to expressions found for similar calculations of nuclear charge radii~\cite{Vanasse:2015fph}.  The trimer radius gives the average distance from the Helium atoms to the Helium trimer c.m. From simple geometry the average distance between Helium atoms, or the bond length, is equivalent to the trimer radius multiplied by $\sqrt{3}$.

Kolganova \emph{et al.}~\cite{Kolganova:2011uc} found a helium-helium scattering length of $a=100.01 \mathrm{\AA}$ and a Helium dimer binding energy of 1.30962 mK using the TTY (Tang, Toennies, and Yiu) potential~\cite{PhysRevLett.74.1546}.  While Roudnev and Yakovlev~\cite{ROUDNEV200097}, using the TTY potential, found a Helium dimer binding energy of 1.312262~mK and did not give a value for the helium-helium scattering length. Using $B_d=\gamma^2/m$ and the binding energy from Roudnev and Yakovlev we find $\gamma=0.01041$ \AA$^{-1}$.  Combining the dimer binding energy from Roudnev and Yakovlev and the scattering length from Kolganova \emph{et al.}, Eq~\eqref{eq:ERErho} can be used to determine an effective range of $r=7.51$~\AA~in good agreement with the value of Ji and Phillips~\cite{Ji:2012nj}, $r=7.50(5)$~\AA.  Fitting the three-body counterterm to reproduce the atom dimer scattering length of 1.205 $\gamma^{-1}$, as determined by Roudnev~\cite{roudnev2003ultra} using the TTY potential, we were able to reproduce the predictions for the ground state, $B_t^{(0)}$, and excited state, $B_t^{(1)}$, trimer energies as found by Ji and Phillips and shown in Table~\ref{tab:results}.  Note, our NLO value for the trimer ground state energy slightly under-predicts the Phillips and Ji value of 89.72 $B_d$.  Changing the effective range to 7.50~\AA~does not alleviate this.  The error of the LO trimer energies are estimated by $\gamma r\sim 8\%$ and the error of the NLO trimer energies by $(\gamma r)^2\sim 0.6\%$.   Comparing to the TTY potential calculations of Roudnev and Yakovlev~\cite{ROUDNEV200097} we find agreement within errors for the excited state trimer energies.  The LO prediction for the ground state trimer energy agrees within error with the TTY prediction~\cite{ROUDNEV200097}, but the NLO prediction does not agree within error.  As noted by Ji and Phillips~\cite{Ji:2012nj} this is likely because the ground state is deeply bound relative to the scale set by the effective range.  The binding momentum of the trimer ground state is roughly $\kappa_3=\sqrt{m B_t^{(0)}}=\gamma\sqrt{96.32}=0.0783$~\AA$^{-1}$.  Multiplying this by the effective range we find an expansion parameter of $\kappa_3r=0.58$, which is considerably larger than the expansion parameter $\gamma r=0.0781$.
\begin{table}[hbt!]
    \centering
    \begin{tabular}{|c|cccc|}\hline
    order  & $B_t^{(1)}$ [$B_d$] & $B_t^{(0)}$ [$B_d$] & $r^{(0)}$ [$\mathrm{\AA}$] & $r^{(1)}$ [$\mathrm{\AA}$]  \\\hline
    LO & 1.723(135) & 97.12(7.59) & 8.35(33) & 103(4)  \\
    NLO  & 1.736(11) & 89.50(55) & 10.29(2) & 105.3(2)  \\\hline
    Exp.~\cite{bruhl2005matter} & & & $11^{+4}_{-5}$ & \\
    Exp.~\cite{Kunitski:2015qth} & & & 10.4 & \\\hline
    TTY Thr.~\cite{ROUDNEV200097} & 1.737 & 96.32 & 10.96 & 105.3 \\\hline
    \end{tabular}
    \caption{\label{tab:results}Binding energies of the ground and excited Efimov trimer at LO and NLO in srEFT.  Their energies are given in units of the Helium dimer binding energy, $B_d$.  Also shown are the LO and NLO srEFT predictions for the bond length of the ground and excited state Efimov trimers.  Various theoretical and experimental determinations of the bond lengths are also shown for comparison.}
\end{table}

Using the same parameters used to calculate the trimer bound state energies we also calculated the average bond length of the ground and excited state trimer as shown in Table~\ref{tab:results}.  Note, due to the square root the error estimate for the bond length is given by $\frac{1}{2}\gamma r\sim 4\%$ ($(\frac{1}{2}\gamma r)^2\sim 0.2\%$) at LO (NLO).  Our NLO prediction for the ground state Helium trimer agrees within errors with the central value from matter wave diffraction of $11^{+4}_{-5}~\mathrm{\AA}$~\cite{bruhl2005matter}.  However, the experimental error is quite large.  Measurements from the Coulomb explosion technique give an average ground state bond length of $10.4~\mathrm{\AA}$~\cite{voigtsberger2014imaging}.  Note, this value did not come with an error likely because the width of the distribution from which it was obtained was larger than the mean value.  The LO and NLO ground state trimer bond lengths disagree within errors with the TTY predictions of Roudnev and Yakovlev~\cite{ROUDNEV200097} shown in Table~\ref{tab:results}.  Again this is likely due to the fact that the ground state energy is smaller than but close to the energy cutoff of our srEFT.  Although data exists for the distribution of bond lengths in the excited state trimer~\cite{Kunitski:2015qth} the average value was not calculated.  Our NLO bond length for the excited state trimer, roughly ten times larger than the ground state trimer, agrees exactly with the TTY prediction of Roudnev and Yakovlev~\cite{ROUDNEV200097}, shown in Table~\ref{tab:results}, and the LO value agrees within its error.

\section{Conclusion\label{sec:Conclusion}}

Cold Helium atoms offer a unique window into Efimov physics.  The weak interaction between Helium atoms naturally gives rise to a single excited Efimov state, without the need for an external magnetic field.  This allows the Coulomb explosion technique to be used to determine the structure of trimer states in Helium and test the predictions of Efimov physics~\cite{Kunitski:2015qth}.  In this work we used srEFT to investigate structural properties of Helium trimers.  Calculating to NLO we included non-universal effects from range corrections.   Given that there is insufficient experimental data and available experimental data has large errors we primarily compared our calculations to theoretical calculations using the TTY potential~\cite{ROUDNEV200097,roudnev2003ultra,Kolganova:2011uc}.  srEFT up to NLO has three parameters that we fit to the dimer binding energy, helium-helium effective range, and atom-dimer scattering length as determined by the TTY potential.  Using these parameters as determined by the TTY potential we were able to reproduce the average bond length of the trimer excited state of TTY potential model calculations.  We found an average ground state trimer bond length of 8.35(33)~$\mathrm{\AA}$ (10.29(2)~$\mathrm{\AA}$) at LO (NLO) and an excited state trimer bond length of 103(4)~$\mathrm{\AA}$  (105.3(2))~$\mathrm{\AA}$ at LO (NLO).  Our calculation of the Helium trimer ground state bond length disagreed with the TTY potential model predictions~\cite{ROUDNEV200097} within errors.  This is further evidence that the srEFT expansion does not work well for the trimer ground state but does work well for the trimer excited state~\cite{Ji:2012nj,Blume:2015nla}.

srEFT can serve as a tool to give correlations between physical observables with the added benefit of allowing for theoretical error estimation.  In principle a NLO srEFT calculation can predict the excited state trimer average bond length to less than 1\%.  However, this is limited by the uncertainty in experimental measurements or theoretical potential model calculations used to fit the parameters in srEFT.  Given the disagreement between potential model calculations and srEFT observables for the trimer ground state it would be prudent to carry this work to higher order in srEFT, as was done  by Ji and Phillips.  A NNLO calculation will be hindered by the introduction of a new three-body counterterm that requires a new three-body datum.  This could in principle be fit to the excited trimer bound state energy as was done by Ji and Phillips~\cite{Ji:2012nj}.  

Future efforts should consider other structural properties of Helium trimers such as bond angles and the distribution of bond lengths.  In this work we used a field theoretic approach to calculate the form factor for the Helium trimer with a fictitious probe that interacted with Helium atoms.  To ease the calculation of structural properties the wavefunctions can be directly obtained from the trimer vertex function in srEFT.  With the wavefunction in hand it is a more straightforward exercise to calculate any structural observable of interest.  Another direction is to consider structural properties of heterogeneous systems in srEFT~\cite{Acharya:2016kjr,Emmons:2017fvb} such as $^3$He$^4$He${_2}$.

\appendix
\section{\label{sec:FFQ2}}

The $Q^2$ contribution to type-a diagrams from Figs.~\ref{fig:LOFF} and~\ref{fig:NLOFF} depends on the values $\mathcal{A}_n$ and functions $\mathcal{A}_n(p)$ and $\mathcal{A}_n(p,k)$ (See Eq.~(\ref{eq:FFQ2A})), where $n=0$ ($n=1$) gives the LO (NLO) contribution. $\mathcal{A}_n$ is given by
\begin{equation}
    \mathcal{A}_n=\int_0^{\Lambda}dqq^2f^{(n)}(q),
\end{equation}
where
\begin{equation}
    f^{(0)}(q)=\frac{m}{576\pi^2}\frac{1}{\Dt^5 D^4}\left(q^2(D^2-2D\Dt+2\Dt^2)+4D\Dt^2(3\Dt-\gamma)\right),
\end{equation}
and
\begin{equation}
    f^{(1)}(q)=\gamma rf^{(0)}(q).
\end{equation}
$\tilde{D}$ and $D$ are defined by
\begin{equation}
    \Dt=\sqrt{\frac{3}{4}q^2-mB_0}\quad,\quad D=\gamma-\Dt.
\end{equation}
The function $\mathcal{A}_n(p)$ is given by
\begin{equation}
    \mathcal{A}_n(p)=\int_0^{\Lambda}dqq^2 f^{(n)}(p,q),
\end{equation}
where
\begin{align}
    &f^{(0)}(p,q)=\frac{4}{3}\bigg[-3\pi f^{(0)}(q)\frac{1}{pq}Q_0(a)\\\nonumber
    &-\frac{m}{27\pi}\frac{1}{D(pq)^3}\left\{\frac{5a}{(1-a^2)^2}+\left[\left(\frac{q}{p}+\frac{p}{q}\right)(1+3a^2)-a(3+a^2)\right]\frac{1}{(1-a^2)^2}\right\}\\\nonumber
    &-\frac{m}{432\pi}\frac{1}{(\Dt D)^3(pq)^2}\left\{\Dt^2D\left[\frac{38}{1-a^2}+\left(\left(20\frac{q}{p}+8\frac{p}{q}\right)a-4(1+a^2)\right)\frac{1}{(1-a^2)^2}\right]\right.\\\nonumber
    &\left.-(\gamma-3\Dt)\frac{9}{2}\frac{q^2}{1-a^2}\right\}\bigg],
\end{align}
\begin{align}
    &f^{(1)}(p,q)=\frac{1}{2}r\left[(\gamma+\Dt)f^{(0)}(p,q)-8\pi Df^{(0)}(q)\frac{1}{pq}Q_0(a)\right.\\\nonumber
    &-\frac{m}{324\pi}\frac{1}{\Dt^3(Dpq)^2}\left\{\left[38\Dt^2D-\frac{9}{2}q^2(\gamma-3\Dt)\right]\frac{1}{1-a^2}\right.\\\nonumber
    &\left.\left.-\Dt^2D\left[4(1+a^2)-\left(20\frac{q}{p}+8\frac{p}{q}\right)a\right]\frac{1}{(1-a^2)^2}\right\}\right],
\end{align}
and,
\begin{equation}
    a=\frac{q^2+p^2-mB_0}{qp}.
\end{equation}
Finally the function $\mathcal{A}_n(p,k)$ is defined by
\begin{equation}
    \mathcal{A}_n(p,k)=\int_0^{\Lambda}dqq^2f^{(n)}(p,k,q),
\end{equation}
where
\begin{align}
    &f^{(0)}(p,k,q)=\frac{16}{3}\bigg[-\frac{6\pi}{8}\left\{f^{(0)}(k,q)\frac{1}{pq}Q_0(a)+f^{(0)}(p,q)\frac{1}{kq}Q_0(b)\right\}\\\nonumber
    &-12\pi^2f^{(0)}(q)\frac{1}{kq}Q_0(b)\frac{1}{pq}Q_0(a)\\\nonumber
    &+\frac{m}{108}\frac{1}{\Dt D^2}\frac{1}{q^4(kp)^2}\left\{2\Dt D\left(\left[12(1-b^2)(1-a^2)+4\frac{q}{p}a(1-b^2)+4\frac{q}{k}b(1-a^2)\right]\right.\right.\\\nonumber
    &+2ab\left[\frac{k}{p}(1-b^2)+\frac{p}{k}(1-a^2)\right]+2b\frac{k}{q}\left[2b^2-(1+a^2)\right]+2a\frac{p}{q}\left[2a^2-(1+b^2)\right]\\\nonumber
    &\left.2\frac{k}{q}\left(\frac{q}{p}a-2\right)(1-b^2)^2Q_0(b)+2\frac{p}{q}\left(\frac{q}{k}b-2\right)(1-a^2)^2Q_0(a)\right)\frac{1}{(1-b^2)^2(1-a^2)^2}\\\nonumber
    &+q^2\left(\left[4+\frac{k}{q}b+\frac{p}{q}a-2\frac{k}{q}\frac{p}{q}ab\right]+\frac{k}{q}(1-b^2)\left(1-2a\frac{p}{q}\right)Q_0(b)\right.\\\nonumber
    &\left.\left.\frac{p}{q}(1-a^2)\left(1-2b\frac{k}{q}\right)Q_0(a)-2\frac{k}{q}\frac{p}{q}(1-b^2)(1-a^2)Q_0(b)Q_0(a)\right)\frac{1}{(1-b^2)(1-a^2)}\right\}\bigg],
\end{align}
\begin{align}
    &f^{(1)}(p,k,q)=\frac{8}{3}\bigg[\frac{3}{16}r(\gamma+\Dt)f^{(0)}(p,k,q)\\\nonumber
    &-\frac{3}{2}\pi f^{(1)}(k,q)\frac{1}{pq}Q_0(a)-\frac{3}{2}\pi f^{(1)}(p,q)\frac{1}{kq}Q_0(b)\\\nonumber
    &+\frac{1}{2}r\frac{m}{54\Dt D(qkp)^2}\left\{\left[4+\frac{k}{q}b+\frac{p}{q}a-2\frac{k}{q}\frac{p}{q}ab\right]\right.\\\nonumber
    &+\frac{k}{q}(1-b^2)\left(1-2a\frac{p}{q}\right)Q_0(b)+\frac{p}{q}(1-a^2)\left(1-2b\frac{k}{q}\right)Q_0(a)\\\nonumber
    &\left.-2\frac{k}{q}\frac{p}{q}(1-b^2)(1-a^2)Q_0(b)Q_0(a)\right\}\frac{1}{(1-b^2)(1-a^2)}\\\nonumber
    &+\frac{3}{2}\pi\frac{1}{2}r(\gamma+\Dt)\left[f^{(0)}(k,q)\frac{1}{pq}Q_0(a)+f^{(0)}(p,q)\frac{1}{kq}Q_0(b)\right]\\\nonumber
    &-6\pi^2\left(f^{(1)}(q)-\frac{1}{2}r(\gamma+\Dt)f^{(0)}(q)\right)\frac{1}{pq}Q_0(a)\frac{1}{kq}Q_0(b)\bigg],
\end{align}
and
\begin{equation}
    b=\frac{q^2+k^2-mB_0}{qk}.
\end{equation}
The $Q^2$ contribution to the form factor from type-b diagrams (See~Eq.~(\ref{eq:FFQ2B})) depends on the function $\mathcal{B}_0(p,k)$ defined by
\begin{equation}
    \mathcal{B}_0(p,k)=\frac{4m\pi}{27}\frac{1}{(pk)^3}\frac{1}{(1-a^2)^2}\left\{\frac{4}{3}\frac{a}{1-a^2}-2a-\frac{1}{3}\frac{p^2+k^2}{pk}\frac{1+3a^2}{1-a^2}\right\}.
\end{equation}
where
\begin{equation}
    a=\frac{p^2+k^2-mB_0}{pk},
\end{equation}
for the rest of the appendix.  $Q^2$ contributions to the form factor from type-c diagrams are given by the functions $\mathcal{C}_n(k)$ and $\mathcal{C}_n(p,k)$ (See Eq.~(\ref{eq:FFQ2C})). $\mathcal{C}_0(k)$ is given by 
\begin{equation}
    \mathcal{C}_0(k)=\frac{m}{576\Dt^5D^3}\left\{4\Dt^2 D(2\Dt-\gamma)+k^2(\gamma-3\Dt)D+2k^2\Dt^2\right\},
\end{equation}
and $\mathcal{C}_1(k)$ by
\begin{equation}
    \mathcal{C}_1(k)=\frac{1}{2}r\left[(\gamma+\Dt)\mathcal{C}_0(k)+\frac{m}{288\Dt^4D^2}\left\{2\Dt^2D+k^2(\Dt-D)\right\}\right],
\end{equation}
where
\begin{equation}
    \tilde{D}=\sqrt{\frac{3}{4}k^2-mB_0}\quad,\quad D=\gamma-\tilde{D},
\end{equation}
for the rest of the appendix.  The function $\mathcal{C}_0(p,k)$ is defined via
\begin{align}
    &\mathcal{C}_0(p,k)=\frac{8}{3}\bigg[-3\pi \mathcal{C}_0(k)\frac{1}{pk}Q_0(a)\\\nonumber
    &-\frac{m\pi}{54\Dt D pk}\left\{\frac{1}{pk}\frac{1}{1-a^2}+\frac{1}{p^2}\left(4a+a\left(\frac{p}{k}\right)^2-2\frac{p}{k}(1+a^2)\right)\frac{1}{(1-a^2)^2}\right\}\\\nonumber
    &-\frac{m\pi}{144}\frac{k}{p}\frac{1}{\Dt^3D^2}\left\{\frac{1}{k^2}Q_0(a)-\frac{1}{pk}\frac{2-\frac{p}{k}a}{1-a^2}\right\}\left[\gamma-3\Dt\right]\bigg],
\end{align}
and $\mathcal{C}_1(p,k)$ by
\begin{align}
    &\mathcal{C}_1(p,k)=\frac{4}{3}r\left[(\gamma+\Dt)\mathcal{C}_0(p,k)\right.\\\nonumber
    &-\frac{m\pi}{96\Dt^4D^2}\frac{1}{pk}Q_0(a)\left\{2\Dt^2D+k^2(\Dt-D)\right\}\\\nonumber
    &\left.-\frac{k}{p}\frac{m\pi}{72\Dt^2D}\left\{\frac{2}{pk}\frac{1}{1-a^2}-\frac{1}{k^2}\frac{1}{1-a^2}-\frac{1}{k^2}Q_0(a)\right\}\right].
\end{align}
Finally, the diagram-d contribution at NLO to the $Q^2$ contribution of the form factor (See Eq.~(\ref{eq:FFQ2D})) depends on the functions
\begin{equation}
    \mathfrak{D}_1(k)=-\frac{m}{576\Dt^3D^3}r\left\{4\Dt^2D+k^2(3\Dt-\gamma)\right\},
\end{equation}
and
\begin{align}
    &\mathfrak{D}_1(p,k)=\frac{8}{3}\bigg[-3\pi\mathfrak{D}_1(k)\frac{1}{pk}Q_0(a)\\\nonumber
    &+\frac{m\pi}{54D(pk)^2}r\left[\left(4\frac{k}{p}+\frac{p}{k}\right)a-3a^2-1\right]\frac{1}{(1-a^2)^2}\\\nonumber
    &-\frac{m\pi}{72\Dt D^2}\frac{1}{pk}r\left\{Q_0(a)+\frac{a-2\frac{k}{p}}{1-a^2}\right\}\bigg].
\end{align}

\end{document}